\documentclass[a4paper,11pt]{article}
\usepackage{pos}

\newcommand{\unit}[1]{\;\mathrm{#1}}

\renewcommand{\printHeadAuthors}{M.~Gintner and J.~Jur\'{a}\v{n}}

\title{The mass exclusion limits for the BSM vector resonances with the direct couplings to the third quark generation}
\ShortTitle{The mass exclusion limits for the BSM vector resonances}

\author[a,c]{Mikul\'{a}\v{s} Gintner}
\author*[a,b]{Josef Jur\'{a}\v{n}}

\affiliation[a]{Institute of Experimental and Applied Physics, Czech Technical University in Prague,\\
Husova 240/5, 110 00 Prague, Czech Republic}

\affiliation[b]{Institute of Physics, Silesian University in Opava,\\
Bezru\v{c}ovo n\'{a}m. 13, 746 01 Opava, Czech Republic}

\affiliation[c]{Physics Department, University of \v{Z}ilina,\\
Univerzitn\'{a} 1, 010 26 \v{Z}ilina, Slovakia}

\emailAdd{gintner@fyzika.uniza.sk}
\emailAdd{josef.juran@utef.cvut.cz}

\abstract{The upper bounds that the LHC measurements searching for heavy resonances beyond the Standard model set on the resonance production cross sections are not universal. They depend on various characteristics of the resonance under consideration, and their validity is also limited by the assumptions and approximations applied to their calculations. The bounds are typically used to derive the mass exclusion limits for the new resonances. 
We address some of the issues that emerge when deriving the mass exclusion limits for the strongly coupled composite $SU(2)_{L+R}$ vector resonance triplet which would interact directly to the third quark generation only.
We show that the presence of such interactions in the model can lower the mass exclusion limits.}

\FullConference{%
  40th International Conference on High Energy physics - ICHEP2020\\
  July 28 - August 6, 2020\\
  Prague, Czech Republic (virtual meeting)
}

\begin{document}
\maketitle

\section{The effective top-BESS model}
\label{sec:tBESS}
One of activities of the ATLAS and CMS Collaborations is searching for new resonances which
are predicted by all major scenarios of the Standard model extension.
Nevertheless,  despite all experimental effort, no new particle has been discovered so 
far. Actually, not even a significant disagreement with the SM predictions has 
been observed yet. This fact has motivated us to calculate mass exclusion limits (MEL)
for the hypothetical top-BESS (tBESS) resonances using the LHC data.

The effective tBESS Lagrangian was formulated and studied in detail in our 
previous papers~\cite{tBESSepjc13,tBESSepjc16,tBESSapp17,tBESSepjc20}. 
It can serve as an 
effective description of the LHC phenomenology of a hypothetical strongly 
interacting extension of the SM where the principal manifestation of this 
scenario would be the existence of a vector resonance triplet as a bound state 
of new strong interactions. The Lagrangian is built to respect the global 
$SU(2)_L\times SU(2)_R\times U(1)_{B-L}\times SU(2)_{HLS}$ symmetry of which the 
$SU(2)_L\times U(1)_Y\times SU(2)_{HLS}$ subgroup is also a local symmetry. The 
$SU(2)_{HLS}$ symmetry is an auxiliary gauge symmetry invoked to accommodate the 
$SU(2)_{L+R}$ triplet of new vector resonances. Each of the mentioned gauge 
groups is accompanied by its gauge coupling: $g$, $g'$, and $g''$ stand for 
$SU(2)_L$, $U(1)_Y$ and $SU(2)_{L+R}$, respectively. Beside the scalar singlet 
representing the 125 GeV Higgs boson and the hypothetical vector triplet, the 
effective Lagrangian is built out of the SM fields only. 

\begin{figure}[htb]
\centerline{
\includegraphics[width=7.3cm]{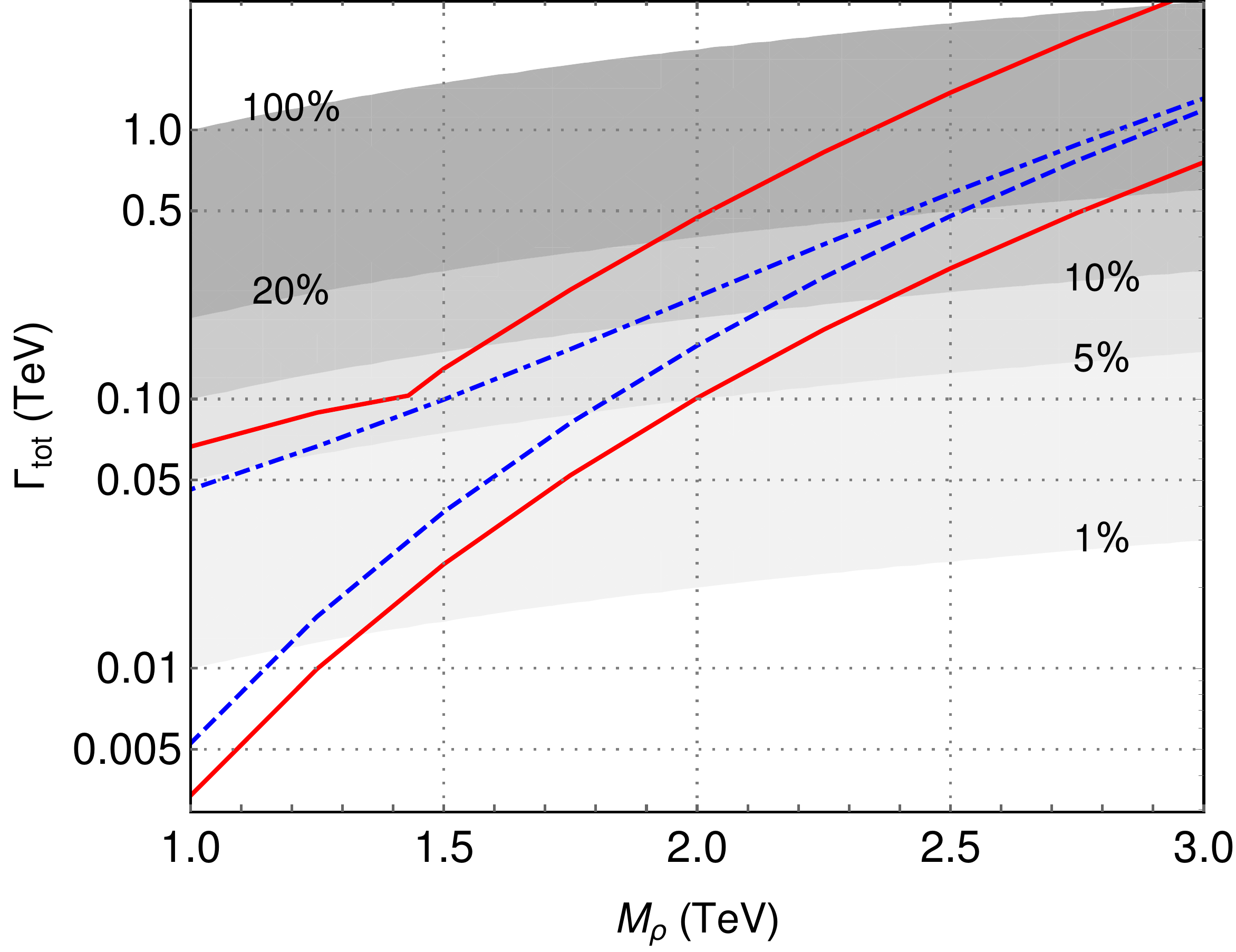}
\hspace{0.4cm}
\includegraphics[width=7.1cm]{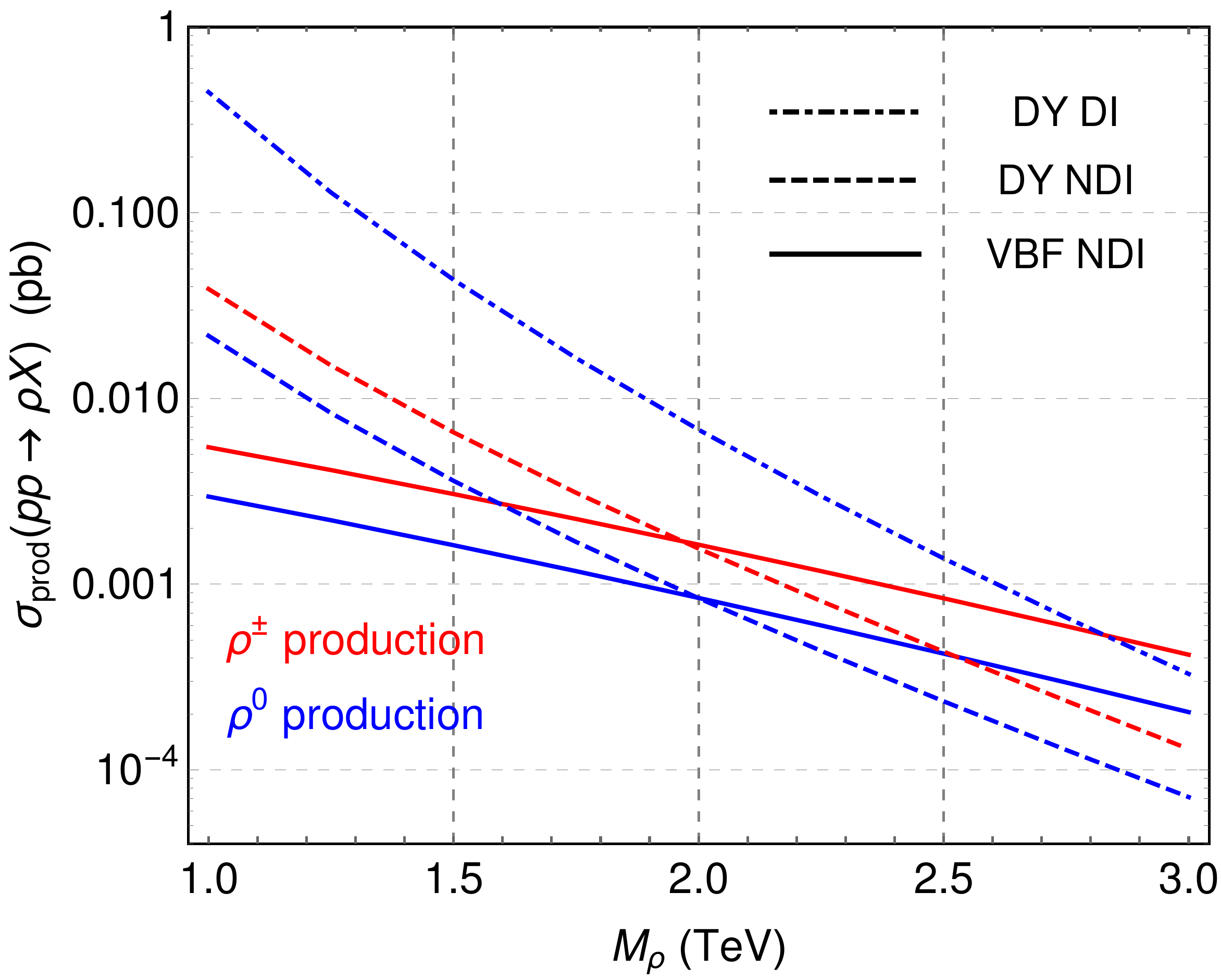}}
\caption{LEFT: The total decay width of the vector resonance,
with no distinction between neutral and charged resonances,
as a function of the resonance mass for $g''=20$: $b_L=b_R=0$ (dashed), and 
$b_{L,R}=-0.1$, $p=1$ (dot-dashed).
The solid curves bound the region of all possible values of the decay width
for the parameter values investigated in the paper. Different shadings 
indicate areas with different intervals of 
$\Gamma_\mathrm{tot}/M_\rho$.\\
RIGHT: The production cross sections $\sigma_\mathrm{prod}(pp\rightarrow\rho 
X)$ for the DY and VBF productions of the neutral (blue) and charged (red) 
vector resonances as functions of the resonance mass when $g''=20$. The solid lines 
stand for the VBF production which is insensitive to $b_{L,R}$. 
The dashed lines denote the DY production when $b_L=b_R=0$. The 
dot-dashed line shows the neutral DY case when
$b_L=b_R=0.1$, $p=1$.}
\label{Fig:TotalDecayWidth}
\end{figure}

There are two specific features of the model that should be called attention to. 
First, the mass and decay widths of the vector resonances are entangled with the model's 
couplings. The vector resonance total width grows quite quickly with the 
resonance mass (see the left panel of Fig.~\ref{Fig:TotalDecayWidth}). 
The masses of the neutral and charged vector resonances are 
virtually degenerate when the resonance coupling $g''$ is much bigger than the 
$SU(2)_L\times U(1)_Y$ gauge couplings $g$ and $g'$.

Secondly, in the fermion sector, the direct interactions (DI) of the 
new vector triplet with the third generation quarks only are considered
in the flavor basis. These interactions grow with $g''$ and are introduced as flavor and chirality dependent. The couplings of the vector resonances to the light SM fermions emerge in the mass eigenstate basis due to the mixing of the gauge bosons. 
These interactions (referred to as the \textit{indirect} ones) are universal and suppressed by $1/g''$. Thanks to the mixing-induced 
couplings, it is possible to produce the vector resonances also in the 
light-quark Drell-Yan processes at the LHC.

In our model, the free pre-factors for the 
couplings to the left and right top-bottom quark doublets are denoted as 
$b_L$ and $b_R$, respectively. 
In the most general case of our model, an additional free pre-factor $p$ is introduced.
It further modifies the direct coupling to the right bottom quark. 
Assuming $0\leq p\leq 1$, the $p$
parameter can be used to suppress the direct coupling to the right bottom quark
relative to the direct coupling to the right top quark.

Following our previous analyses~\cite{tBESSepjc13,tBESSepjc16}
we restrict our MEL searches to the following region of the 
parameter space: 
$1\leq M_\rho/\mathrm{TeV}\leq 3$, $12\leq g''\leq 25$, 
$|b_{L,R}|\leq 0.1$, and $0\leq p \leq 1$.
These restrictions are based on various experimental and theoretical considerations
such as the low-energy analysis (EWPT), unitarity limits, etc.


\section{The mass exclusion limits}
\label{sec:MELs}

We have obtained the mass exclusion limits on the tBESS vector resonances
from the experimental upper bounds on the $s$-channel resonance 
production cross section times branching ratio provided by the ATLAS and CMS 
Collaborations which were derived with the narrow resonance qualification. 
Consistently, the model cross section predictions have 
been calculated in the narrow width approximation (NWA), ie.
$\sigma(pp\rightarrow abX)= 
        \sigma_\mathrm{prod}(pp\rightarrow \rho X)\times\mathrm{BR}(\rho\rightarrow ab)$,
where $\sigma_\mathrm{prod}$ is the on-shell cross section for the vector 
resonance production, 
and $\mathrm{BR}(\rho\rightarrow ab)$ is the branching ratio for 
the vector resonance decay channel.
It is generally expected that the approximation works 
well when $\Gamma_\mathrm{tot}/M_\rho\lesssim 10\%$.
The NWA also ignores the signal-background 
interference effects. Their influence on the precision of the 
NWA have been inspected in~\cite{Pappadopulo_etal14}.

We approximate $\sigma_\mathrm{prod}$ by the sum of the 
vector boson fusion (VBF) and the Drell-Yan (DY) cross sections. 
In the VBF production, the vector resonance can emerge from $W^+W^-\rightarrow\rho^0$
and $W^+ Z\rightarrow \rho^+$ (+c.c.). This production is calculated in
the Effective-W Approximation~\cite{Dawson1985EWA} with the longitudinal 
W/Z degrees of freedom only.
The DY production of our triplet can proceed via  $u\bar{u}, d\bar{d}, 
c\bar{c}, s\bar{s}, b\bar{b}\rightarrow \rho^0$, and $u\bar{d}, 
c\bar{s}\rightarrow \rho^+$ (+c.c.).
We have observed that the $b$-quark partonic contents of protons cannot be ignored
for non-zero DI's.  
The production of the neutral resonance via $b\bar{b}$ annihilation can dramatically affect
$\sigma_\mathrm{prod}$.
There are parameter space regions where over $90\%$ of the neutral resonance 
production is comprised of $b\bar{b}\rightarrow \rho^0$. Consequently, the 
sensitivity of $b\bar{b}\rightarrow \rho^0$ to the DI couplings 
impacts all MEL's based on the neutral resonance channels.
This effect is illustrated in the right panel of Fig.~\ref{Fig:TotalDecayWidth}.
As can also be seen there both production mechanisms contribute comparably
in the range of investigated masses.

The second factor of the NWA formula is the branching ratio of the decay channel. 
In our model, if there is no direct interaction (NDI) the vector triplet decays mainly to the electroweak gauge bosons,
$\mathrm{BR}(WW,WZ)\approx 1$. With non-zero DI's the decay channels to the top and bottom quarks can compete with $WZ$ and $WW$ ones. The other channels are negligible.

The regions of the parameter space where the predicted cross section exceeds
the experimental upper bounds are excluded for a given model.
We have reviewed fourteen vector resonance decay channels available to the date
of this analysis:
$WZ$,
$WW$,
$WH$, 
$ZH$, 
$jj$, 
$\ell\ell$, 
$\ell\nu$, 
$\tau\tau$, 
$\tau\nu$, 
$tt$,
$bb$,
and 
$tb$, where $\ell=e,\mu$. 
The related $95\%$~C.L.\ upper bounds
are based on the integrated luminosity of $13.2-139\unit{fb}^{-1}$ 
of proton-proton collision data at $\sqrt{s}=13\unit{TeV}$ measured by ATLAS and CMS Collaborations.

\begin{figure*}[thb]
\centerline{
\includegraphics[width=7.3cm]{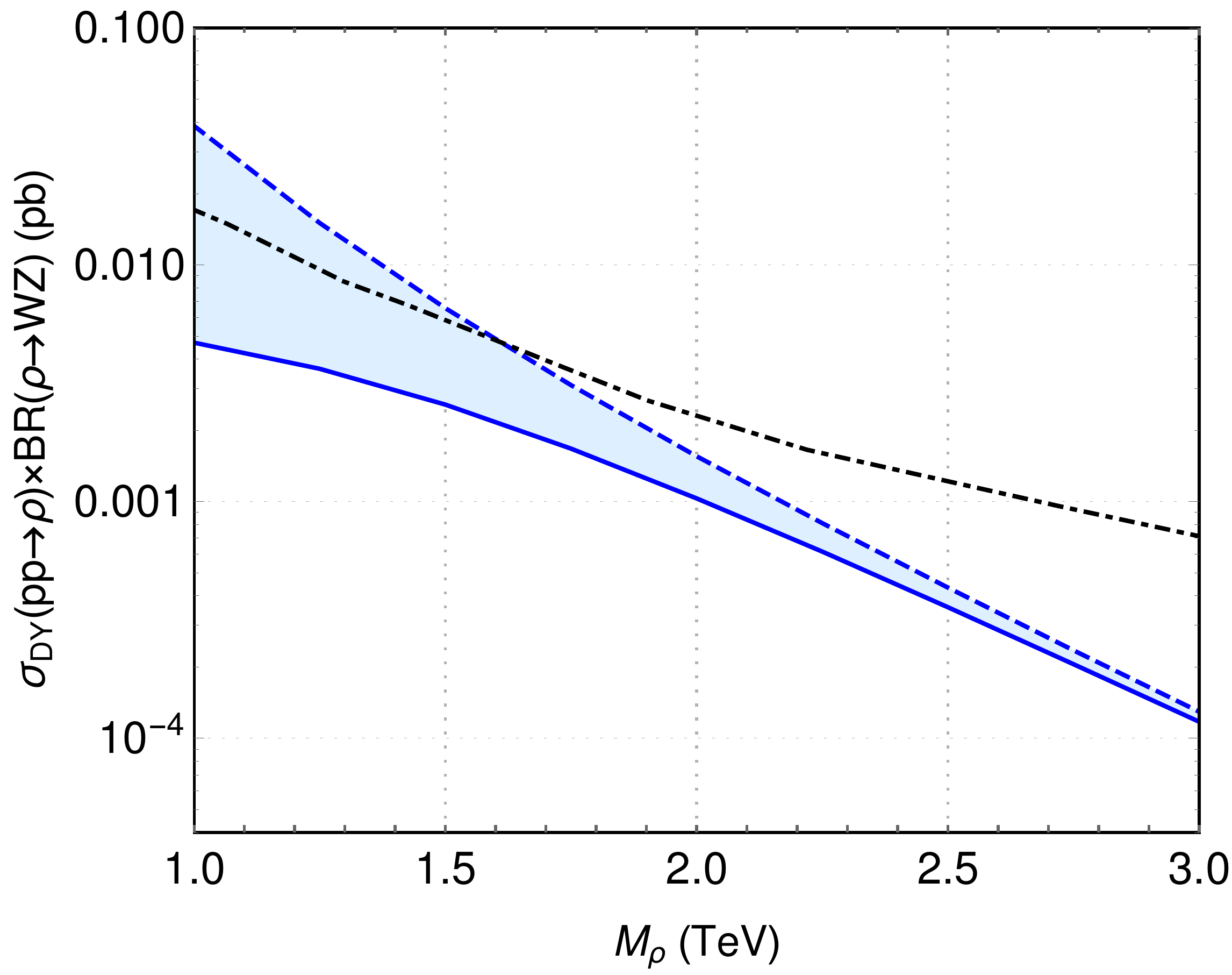}
\hspace{0.4cm}
\includegraphics[width=7.3cm]{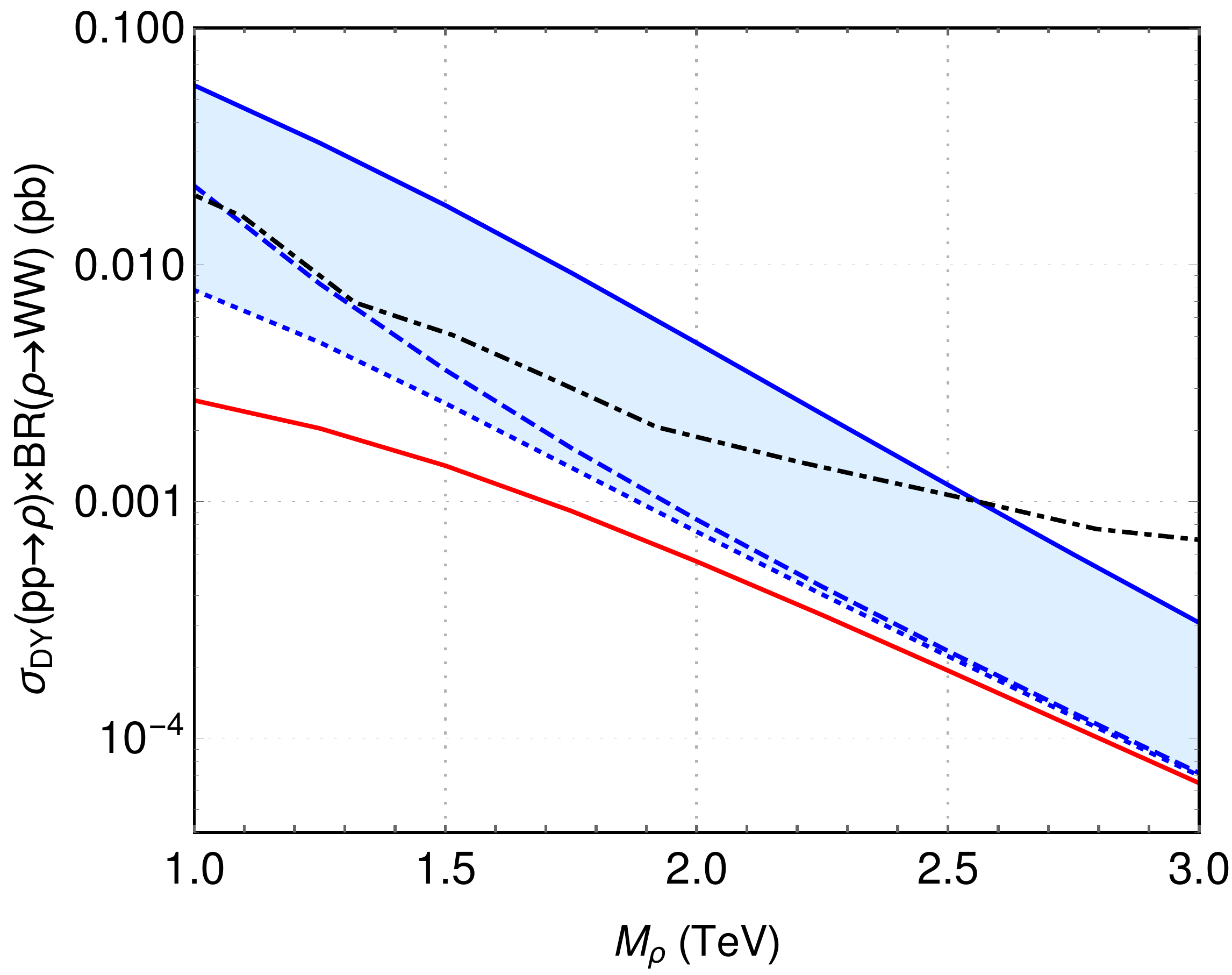}}
\caption{The bands of the possible values of $\sigma^\mathrm{prod}_{DY}\times\mathrm{BR}$
for the tBESS model DI couplings under consideration (ligth blue areas) 
for the $WZ$ (left) and $WW$ (right) channels as functions of $M_\rho$ for $g''=20$. 
The dot-dashed black lines indicate 
the relevant experimental upper bounds~\cite{P5}.
The curves of selected cross sections are also shown:
$b_{L,R}=0$ (dashed blue),
$b_L=-0.1,b_R=0.1,p=1$ (the solid blue lines when the bottom quark partons are included,
the solid red line without the bottom quark partons), and
$b_L=0,b_R=-0.1,p=0$ (dotted blue curve).
}
\label{Fig:MEL}
\end{figure*}

Of all investigated processes the channels with $WW$ and $WZ$ in the final state only
provide the MEL's for the tBESS vector triplet.
Regarding the top and bottom quark final states
the integrated luminosity of the order $1\unit{ab}^{-1}$ would 
be needed to provide for any exclusion limits at all.
To see the impact of the DI couplings on the resulting MEL
we have investigated the cross section dependency on parameters of $b_L,b_R,p$. 
Of course, the non-zero DI couplings make the analysis of 
the limits significantly more complex problem.

All important channels follow one of two qualitatively different behaviors.
In the first case, the $WZ_\mathrm{DY}$, $WZ_\mathrm{VBF}$, 
$WZ_\mathrm{DY+VBF}$, and $WW_\mathrm{VBF}$ cross sections diminish in 
comparison with the NDI case for all parameter values under 
consideration. It is because these cross sections depend on $b_{L,R}$ and $p$ 
through $\mathrm{BR}(\rho\rightarrow WW)$ and $\mathrm{BR}(\rho\rightarrow WZ)$ 
only. When $|b_{L,R}|\leq 0.1$ the cross sections for all four modes
lie in the bands between the NDI cross section (the upper 
boundary) and
the $|b_L|=|b_R|=0.1$ and $p=1$ cross section (the lower boundary).
In the later case, the cross sections virtually do not depend on the signs of $b_{L,R}$.
As an example, in Fig.~\ref{Fig:MEL} (left), we show 
the process of the DY production of the charged resonance 
followed by its decay to $W$ and $Z$ bosons.
Cross sections for all considered values of the DI couplings lie within the light blue area.
We can see that there are combinations of DI's for which the vector resonances of the mass
below about $1.6\unit{TeV}$ are excluded. However, there are also regions of the parameter space where all masses under consideration are admissible.

In the second case, the behavior of the $WW_\mathrm{DY}$ and $WW_\mathrm{DY+VBF}$
cross sections is not so easy to conjecture. 
Its sensitivity to the direct interactions originates not only in 
$\mathrm{BR}(\rho\rightarrow WW)$ but also in the production of $\rho$ through 
the $b\bar{b}$ annihilation. Their cross 
sections can either grow or decrease with the growing strength of the DI's, 
depending on the particular combination of the $(b_L,b_R,p)$ 
parameter values. This more complex behavior originates from the competition 
between the growing production cross section and the shrinking branching ratio. 
The resulting cross sections are bound from above by the cross section 
for $b_L=-b_R=-0.1,p=1$ and from below for parameters $b_L=0, b_R=-0.1, p=0$.
As an example, we can see in Fig.~\ref{Fig:MEL} (right) cross sections
of DY production of neutral resonance followed by decay to $WW$ bosons. 
All possible values of the cross sections lie between the upper (solid blue line)
and lower (dotted blue line) boundaries.

In Fig.~\ref{Fig:MEL} (right), we also demonstrate the impact of the $b$-quark 
partonic contents in the colliding protons.
It affects only the production of neutral resonance via $b\bar{b}$ annihilation.
If we consider $WW_\mathrm{DY}$ channel with $b_L=-0.1, b_R=0.1, p=1$ 
we obtain prediction described
by the solid blue line. Neglecting the $b$ quark partons
would result in the cross section represented by the the solid red line.
With respect to the NDI case, the non-zero DI couplings increase the cross section 
when the bottom quark partons are considered and they decrease the cross section when the bottom quark partons are neglected.

The resulting MEL comes from the most stringent limit of all channels. 
We compare limits from the following channels:
$WZ_\mathrm{DY}$, $WZ_\mathrm{DY+VBF}$, $WW_\mathrm{DY}$, and $WW_\mathrm{DY+VBF}$.
Due to the mass degeneracy of the neutral and charged vector resonances 
both resonances are excluded by the higher of the two limits.
In the case of NDI the resulting MEL's are summarized in Table~\ref{Tab:MEL} in the third row. 
The exclusion limits for $g''$ values below $20$ are not shown because their fatness 
(shown in the second row of the Table)
exceeds $40\%$ which makes the limits obtained via the NWA calculations unreliable.
In the case of DI's we need some additional restrictions on the free parameters $b_L, b_R, p$ 
to obtain MEL. In the following we assume two DI scenarios: 
i) $b\equiv b_L=b_R$ is free and $p=1$ and 
ii) $b_L=0$, $b_R=0.1$ and $p$ is free.
In both cases the DI is parameterized by a single free parameter. 
In the former case the resulting MEL for given $g''$ is done by the smallest MEL 
when $|b|\leq 0.1$ and in the latter case when $0\leq p\leq 1$.
Results are summarized in the Table~\ref{Tab:MEL} in the forth and the fifth rows, respectively.

\begin{table*}[htb]
\centering
\caption{The (smallest) MEL in TeV for the tBESS vector resonance triplet 
for various values of $g''$ in the case of NDI,
and for specific DI scenarios:
i) assuming $p=1$ and $b\equiv b_L=b_R$ within the interval $|b|\leq 0.1$;
ii) assuming $b_L=0$, $b_R=0.1$, and within the interval $0\leq p\leq 1$.
The second row shows the resonance fatness for NDI case (which values are also closed to DI cases).}
\label{Tab:MEL}
\begin{tabular}{c|ccccccccccc}
\hline
$g''$ & \multicolumn{5}{c}{12 -- 19} & 20 & 21 & 22 & 23 & 24 & 25\\
\hline
$\Gamma_\mathrm{tot}/M_\rho$ 
& \multicolumn{5}{c}{$> 0.44$} & 0.40 & 0.33 & 0.26 & 0.21 & 0.17 & 0.13 \\
NDI & \multicolumn{5}{c}{$> 3$} & $> 3$ & 2.94 & 2.84 & 2.73 & 2.65 & 2.53$^a$\\
\hline
DI scenario i) & \multicolumn{5}{c}{$> 3$} & 2.96 & 2.86 & 2.74 & 2.65 & 2.52$^a$ & 2.41$^a$\\
DI scenario ii) & \multicolumn{5}{c}{$> 3$} & 2.97 & 2.87 & 2.73 & 2.63 & 2.45 & 2.28\\
\hline
\end{tabular}\\
$^a$ the mass is also not excluded in a small interval below $1.3\unit{TeV}$
\end{table*}

To conclude, the bottom quark partonic contents can provide an important contribution to the neutral resonance production in the DY mode.
Its disregard would alter behaviour of some channels qualitatively.
The considered values of DI's can decrease the MEL's up to $10\%$ 
with respect to the NDI case\footnote{There are exceptions where small intervals
of unrestricted masses below $1.3\unit{TeV}$ exist.
However, this takes place for the values of $g''$ which are close to the naive perturbativity limit of 25.}.
The obtained MEL's in the paper exceed $2.28\unit{TeV}$ where
the width-to-mass ratio is higher than $11\%$. 
Should the MEL's exceed $\approx 3\unit{TeV}$ a thorougher analysis beyond 
the NWA is needed ~\cite{Xie2019}.
This would require an additional effort on the side of data simulations and analysis along with new phenomenology calculations.


\acknowledgments
The work was supported by the grant LTT17018 of the Ministry of Education, 
Youth and Sports of the Czech Republic and by the COST Action CA15108
"Connecting insights in fundamental physics".
M.G.\ was supported by the Slovak CERN Fund. We would also like to thank the
Slovak Institute for Basic Research for their support.




\begin{thebibliography}{99}



\bibitem{tBESSepjc13}
M.~Gintner, J.~Jur\'{a}\v{n},
\emph{The vector resonance triplet with the direct coupling to the third quark generation},
\href{https://doi.org/10.1140/epjc/s10052-013-2577-5}{\emph{Eur. Phys. J. C} \textbf{73} (2013) 2577}
[{\tt hep-ph/1309.6597}].

\bibitem{tBESSepjc16}
M.~Gintner, J.~Jur\'{a}\v{n},
\emph{The limits on the strong Higgs sector parameters in the presence of new vector resonances},
\href{https://doi.org/10.1140/epjc/s10052-016-4484-z}{\emph{Eur. Phys. J. C} \textbf{76} (2016) 651}
[{\tt hep-ph/1608.00463}];
\emph{Erratum} 
\href{https://doi.org/10.1140/epjc/s10052-016-4579-6}{\emph{Eur. Phys. J. C} \textbf{77} (2017) 6}.

\bibitem{tBESSapp17}
M.~Gintner, J.~Jur\'{a}\v{n},
\emph{The LHC mass limits for the SU(2)L+R vector resonance triplet of a strong extension of the Standard model},
\href{https://doi.org/10.5506/APhysPolB.48.1383}{\emph{Acta Phys. Pol. B} \textbf{48} (2017) 1383}
[{\tt hep-ph/1705.04806}].

\bibitem{tBESSepjc20}
M.~Gintner, J.~Jur\'{a}\v{n},
\emph{A case study about the mass exclusion limits for the BSM vector resonances with the direct couplings to the third quark generation},
\href{https://doi.org/10.1140/epjc/s10052-020-7732-1}{\emph{Eur. Phys. J. C} \textbf{80} (2020) 161}
[{\tt hep-ph/1908.11619}].

\bibitem{Pappadopulo_etal14}
D. Pappadopulo, A. Thamm, R. Torre and A. Wulzer,
\emph{Heavy Vector Triplets: Bridging Theory and Data},
\href{https://doi.org/10.1007/JHEP09(2014)060}{\emph{JHEP} \textbf{1409} (2014) 060}
[{\tt hep-ph/1402.4431}].

\bibitem{Dawson1985EWA}
S. Dawson,
\emph{The effective W approximation},
\href{https://doi.org/10.1016/0550-3213(85)90038-0}{\emph{Nucl. Phys.} \textbf{B249} (1985) 42}.

\bibitem{P5}
ATLAS Collaboration,
\emph{Combination of searches for heavy resonances decaying into bosonic and leptonic final states using 36 fb$^{-1}$ of proton-proton collision data at $\sqrt{s}$=13 TeV with the ATLAS detector},
\href{https://atlas.web.cern.ch/Atlas/GROUPS/PHYSICS/PAPERS/EXOT-2017-31/}{\emph{ATLAS-CONF-EXOT-2017-31}.}

\bibitem{Xie2019}
D. Liu, L.-T. Wang, and K.-P. Xie,
\emph{Broad composite resonances and their signals at the LHC},
\href{https://doi.org/10.1103/PhysRevD.100.075021}{\emph{Phys. Rev. D} \textbf{100} (2019) 075021}
[{\tt hep-ph/1901.01674}].

\end{thebibliography}
\end{document}